\documentclass[preprint,1p]{elsarticle}
\usepackage[usenames, dvipsnames]{xcolor}
\usepackage{geometry}
\usepackage[utf8]{inputenc}
\usepackage[T1]{fontenc}
\usepackage{amsmath}
\usepackage{amssymb}
\usepackage{graphicx}
\usepackage{multicol}
\usepackage{multirow}
\usepackage{longtable}
\usepackage{pbox}
\usepackage{array,multirow,graphicx}
\usepackage{natbib}
\usepackage[version=3]{mhchem}
\usepackage{footnote}
\usepackage[utf8]{inputenc}
\usepackage{scrextend}
\usepackage{diagbox}
\usepackage{hyperref}
\usepackage{MnSymbol}
\usepackage{tikz}
\usepackage{ctable}

\usepackage[titletoc,toc,title]{appendix}

\DeclareRobustCommand{\filledcirc}[1]{%
  \mathrel{%
    \tikz{%
      \draw[black,fill=#1] (0,0) circle (.5ex);%
    }%
  }%
}

\newcommand{\specialcell}[2][c]{%
  \begin{tabular}[#1]{@{}l@{}}#2\end{tabular}}

\author[sil,cebi]{Katarzyna Jesionek}
\author[sil,cebi]{Aleksandra Slapik}
\author[sil,cebi]{ Marcin Kostur}
\address[sil]{Institute of Physics, University of Silesia, 40-007 Katowice, Poland}
\address[cebi]{Silesian Center for Education and Interdisciplinary Research, University of Silesia, 41-500 Chorzów, Poland}

\date{\today}

\begin{document}
%------------------------------------------------

%--BEGIN FRONTMATTER-----------------------------
\begin{frontmatter}

%--TITLE-----------------------------------------
\title{Effect of hypertension on low-density lipoprotein transport within a multi-layered arterial wall: modelling consistent with experiments}
%--ABSTRACT--------------------------------------
\begin{abstract}

The influence of hypertension on low-density lipoproteins intake into the arterial wall is an important factor for understanding  mechanisms of atherosclerosis. It has been experimentally observed that the increased pressure leads to the higher level of the LDL inside the wall.  In this paper we attempt to construct a model of the LDL transport which reproduces quantitatively experimental outcomes. We supplement the well known four-layer arterial wall model to include two pressure induced effects: the compression of the intima tissue and the increase of the fraction of leaky junctions.  We demonstrate that such model can reach the very good agreement with experimental data. 

\end{abstract}

%--KEYWORDS--------------------------------------

\end{frontmatter}
%--END FRONTMATTER-------------------------------

%------------------------------------------------
%------------------------------------------------
\normalsize
\section{Introduction}

Cardiovascular diseases are among the most serious problems of the modern civilization. It has been estimated that in developed societies more than the half of all deaths ascertained every year is caused by atherosclerosis \cite{chugh2008epidemiology, rosenberg2014height, chugh2004current}. Unfortunately, the pathogenesis of this disease is still not fully understood. However, it is generally accepted that the development of atherosclerotic plaque is associated with low-density lipoproteins (LDL), which are responsible for the distribution of cholesterol and triacylglycerides among the body cells \cite{mahley1988apolipoprotein}. According to the widely conducted research, it is commonly accepted that atherosclerotic lesions are preceded by the abnormally high concentration of the LDL in the intima \cite{chobanian1989influence, tarbell2003mass, stary1988evolution, meyer}. Moreover, the experimental, epidemiological, and postmortem studies have revealed that development of this disease is enhanced by the hypertension and low or oscillatory  wall shear stress (WSS) \cite{chobanian1989influence, meyer, glagov1988hemodynamics, yiannis2007role}. Investigation of mechanisms of the LDL macromolecules transport in the arterial wall affected by these factors is therefore a very important step leading toward the understanding of atherosclerosis pathogenesis. 

It has been experimentally observed that the hypertension causes the increase in the relative LDL concentration in the arterial wall \cite{chobanian1989influence, meyer}. It used to be qualitatively explained by the  enhanced endothelium permeability and increased water filtration through the wall \cite{curmi1990effect,bratzler}. Moreover, it has been reported that the mechanical compression of the intima  under increased pressure could be an important factor \cite{huang}.

The LDL transport has also been extensively subjected to theoretical modeling \cite{olgac2008computational, sun2007effects, giugliano1995diabetes, dabagh2009transport, ai_vafai_2006, khakpour2008effects}. In several studies the impact of the pressure on the LDL transport was investigated \cite{dabagh2009transport, vafai_yang_2006, liu2011effect}. In general, since models are based on the filtration process, they all include the increased water filtration due to the hypertension. However, for high pressure this effect alone predicts much smaller LDL concentration inside the vessel wall (see for example in  \cite{vafai_yang_2006, wang2015analysis}) than observed experimentally \cite{meyer, curmi1990effect}. It suggests that other pressure induced effects should be included in the LDL transport models. We propose to supplement these models by mechanical compression of the intima and pressure induced changes in endothelium structure.

The compression of intima has been extensively studied in \cite{huang} and thus we calculate the thickness of the compressed intima using data from that work. This effect has been already taken into account in theoretical study by Dabagh et al. in \cite{dabagh2009transport} and Liu et al. in \cite{liu2011effect}.  The impact of the transmural pressure on the endothelial fraction of leaky junctions is, according to the best of authors' knowledge, not directly known and there are different values used in the literature \cite{dabagh2009transport, liu2011effect, wu1990transendothelial}. Moreover, Dabagh et al. in \cite{dabagh2009transport} suggested that endothelium reacts to the elevated pressure with some delay. It means that the fraction of leaky junctions increases in the first hour of experiment under hypertension. 

In this paper we introduce LDL transport model, which takes into account all three mentioned effects of the elevated transmural pressure: increase of the filtration velocity, pressure and time dependent increase of the endothelial fraction of leaky junctions and the compression of intima. In our model we want to reconstruct  biological conditions and reproduce experimental results as accurate as possible. Because there is no experimental or theoretical quantitative relation between the fraction of leaky junctions and hypertension, we investigate the dependence of this fraction on the pressure and also on time of the exposure to hypertension. Based on available experimentally obtained LDL concentration profiles in the arterial wall, we obtain stationary values of the fraction of leaky junctions for two levels of hypertension ($120 \, \mathrm{mmHg}$ and $160 \, \mathrm{mmHg}$) and the relationship between the fraction of leaky junctions and the time of exposition to the elevated pressure. In this purpose we fit LDL mean tissue concentration profiles to available experimental results for different values of pressure and incubation time.

Since we need to perform a large amount of computationally demanding calculations, we decided to reduce numerical complexity of the model by reduction of its dimensionality. The blood vessel is clearly a three-dimensional system. However, processes of permeation through the vessel wall are dominated by the pressure forced filtration, and predominantly occur in the radial direction. Hence, the process of LDL transport takes place essentially in one direction: from the inside of the vessel wall to the outside. Because of this fact, in this paper we consider one-dimensional model.  A comparison of two- and three-dimensional models with one-dimensional models presented in the literature confirms legitimacy of this simplification \cite{khakpour2008effects, our, michel1999, moore1997oxygen}. 

%------------------------------------------------ 
%------------------------------------------------
\section{Multi-layer model of the arterial wall}
\label{sec:model}

\begin{figure}[t] 
\centering
\includegraphics[width=0.7\linewidth]{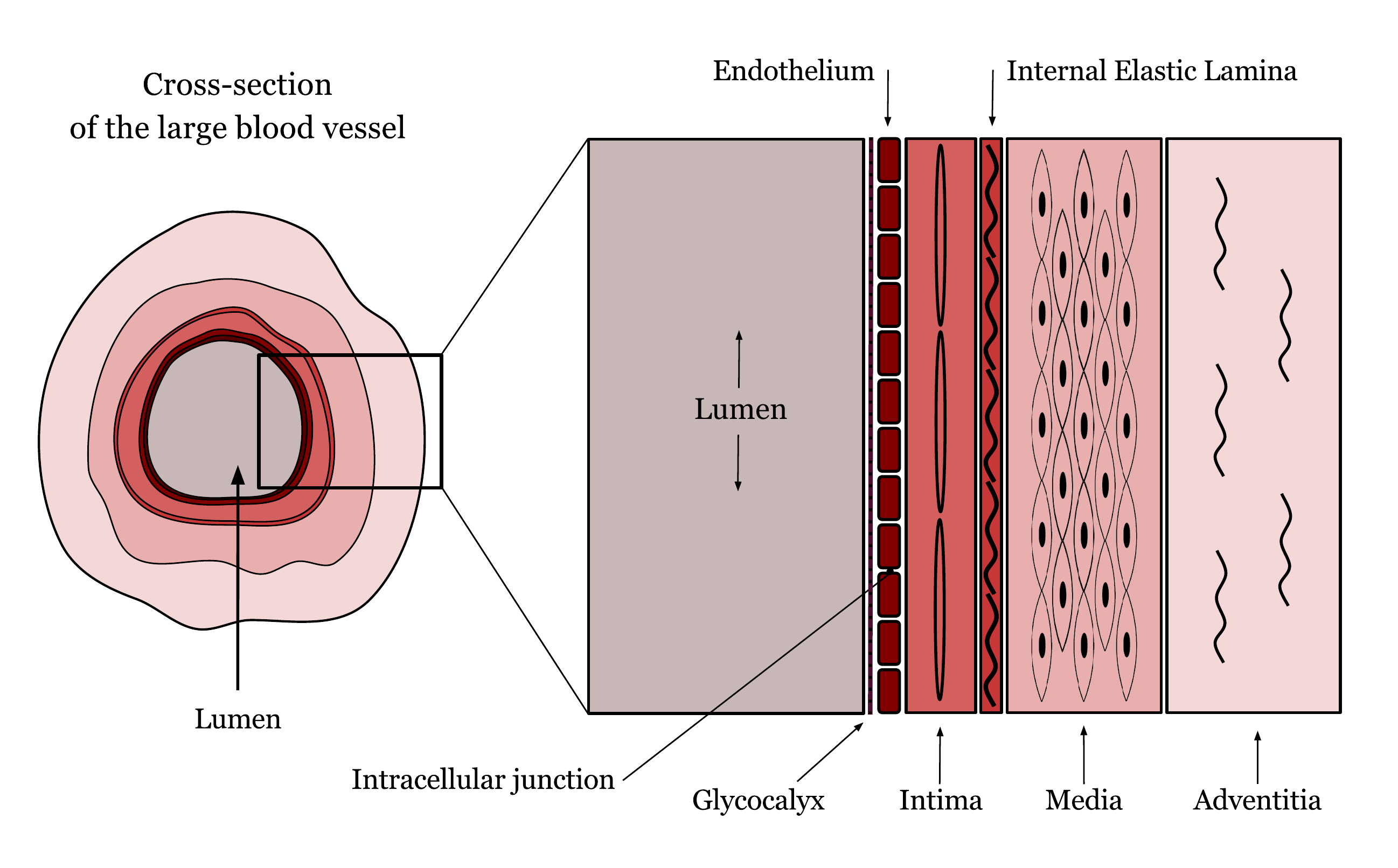}
\caption{The layered structure of the arterial wall. }
\label{fig:StructBV}
\end{figure}

A typical large blood arterial wall has a layered structure, schematically shown in Figure \ref{fig:StructBV}. Going from the lumen, the large artery consists of following six layers: glycocalyx, endothelium, intima, internal elastic lamina (IEL), media and adventitia \cite{vafai_yang_2006}. Such layered structure can induce some important effects and thus it seems reasonable to include it in computer modeling. 

The multi-layer model was first proposed in \cite{ai_vafai_2006} by Ai and Vafai. It covers four major arterial wall layers: endothelium, intima, IEL and media, which are treated as homogeneous materials with different parameters. In this model the impact of the glycocalyx is neglected due to its very small thickness and due to the fact that it does not obstruct the main LDL entries - leaky junctions. Moreover, adventitia is treated as a boundary condition.

The transport of LDL macromolecules in the layered arterial wall is complex and multiscale process. In order to obtain a numerically feasible model, we consider an effective system consisting of two membranes: endothelium and IEL, and two porous layers: intima and media. Membranes are modeled using filtration theory, where osmosis can be neglected \cite{ai_vafai_2006, vafai_yang_2006}. Therefore, in our approach all layers can be treated as macroscopically homogeneous porous media \cite{ai_vafai_2006, vafai_yang_2006}.  The LDL transport in each layer is thus mathematically modeled using volume averaged porous media equations, with Staverman filtration reflection coefficient:
\begin{eqnarray}
\label{eq:velocity}
	\nabla \cdot \vec{u} & = & 0, \\
\label{eq:pde1}
      \frac{\rho}{\epsilon_i} \frac{\partial \vec{u}}{\partial t} + \frac{\mu}{K_i}\vec{u} & = & - \nabla P + \frac{\mu}{\epsilon_i} \nabla^2 \vec{u}, \\
\label{eq:pde2}
      \epsilon_i\frac{\partial c}{\partial t} +(1-\sigma_i)\vec u\cdot\nabla c & = & D^{eff}_i \nabla^2 c - k_i c, 
\end{eqnarray}
where
$\vec u$ is the volume averaged filtration velocity of the solvent penetration, $P$ is the pressure, $c$ is the dimensionless LDL concentration in the fluid phase normalized to the input (i.e. lumen) concentration, $\mu$ is the fluid dynamic viscosity, $K_i$ is the permeability of the i-th layer, $\rho$ is the fluid density, $\epsilon_i$ is the porosity of the i-th layer, $\sigma_i$ is the LDL reflection coefficient in the i-th layer, $D^{eff}_i$ is the LDL effective diffusivity of the i-th layer and $t$ denotes time. The model assumes first order decay of the LDL with the reaction rate coefficient $k_i$ in the i-th layer. The coefficient $k_i$ is equal to $0$ everywhere except the media layer, where we denote it as $k_4=k$.

%------------------------------------------------
\subsection{Endothelium}

Endothelium is a membrane in the direct contact with blood. It is the most sensitive to hemodynamical conditions in the blood vessel, such as a blood pressure or a wall shear stress. This layer makes the largest contribution to the hydraulic and the mass transfer resistance in the vessel wall. Therefore, factors causing increase in the amount of pores and in their effective size have a significant impact on the transport in the entire wall \cite{olgac2008computational, our}. 

Endothelium cells are connected to each other by so called intracellular junctions. These junctions would normally not allow for any significant passage of LDL molecules, even through the breaks inside normal junctions, because the wide part of the cleft is expected to be of the order of LDL molecule size. The average radius of the normal junction is $5.5 \, \mbox{nm}$, which is smaller than the radius of the LDL molecule ($r_{m} =11 \, \mbox{nm}$)  \cite{tarbell2003mass, our}.   

\begin{figure}[t] 
	\centering \includegraphics [width=0.6\textwidth]{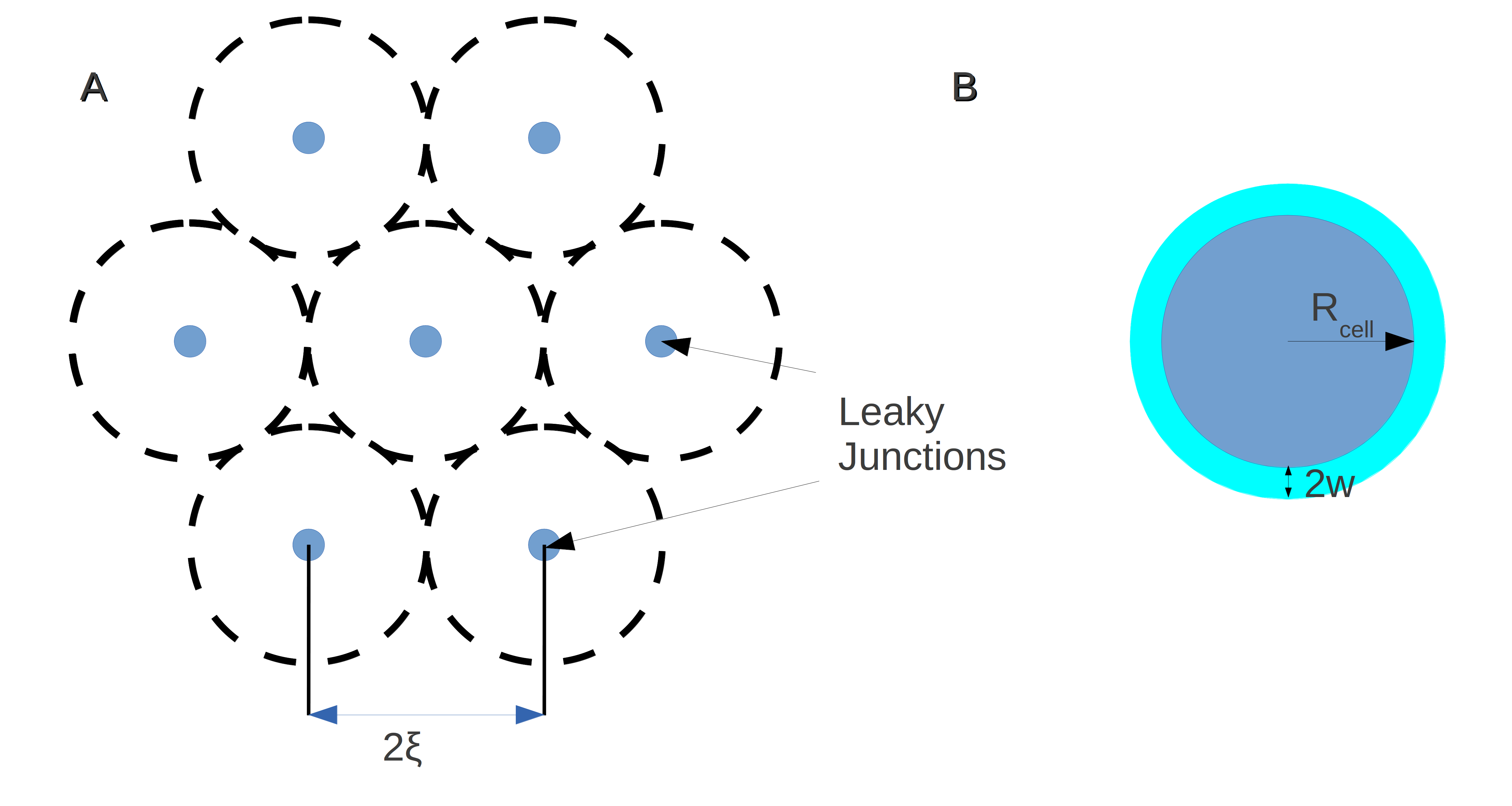}
	\caption{Randomly distributed leaky cells \cite{our}. (A)
          A leaky cell is illustrated by the blue circle in the
          center of each periodic circular unit of radius $\xi$,
          represented by dashed lines. (B) Leaky cell of radius
          $R_{cell}$ with leaky junction of half-width $w = 14.3 \, \mbox{nm}$.}
	\label{fig:leaky}
\end{figure}

However, intracellular junctions can be leaky. These large, leaky pores of the width between $20 \, \mathrm{nm}$ and $40 \, \mathrm{nm}$ \cite{olgac2008computational, formaggia2010cardiovascular} are associated with cells that are in the process of cell turnover: either cell division (mitosis) or cell death (apoptosis) \cite{tarbell2003mass, vafai2012}. They are formed up due to the weakening of cells junctions during the process of division or sloughing off cells by healthy neighbors \cite{tarbell2003mass}. The number of leaky cells is expressed by the fraction of leaky junctions defined as a ratio of the leaky cells area to the area of all cells \cite{dabagh2009transport}
\begin{eqnarray}
	\label{eq:PHI}
    \phi=\frac{R_{cell}^{2}}{\xi^{2}},
\end{eqnarray}
where $R_{cell}$ is the radius of endothelial cell ($R_{cell} = 15 \, \mu \mbox{m}$) and $\xi$ is the radius of periodic circular unit dependent on the number of leaky junctions \cite{huang1994fiber}. It is shown schematically in Figure \ref{fig:leaky}.

It turns out that processes of apoptosis and mitosis are affected not only by the wall shear stress, as mentioned previously in \cite{our}, but also by the pressure \cite{curmi1990effect, dabagh2009transport, bretherton1976effect}. Therefore, in our model, the hypertension increases the fraction of leaky junctions and influences all endothelial transport parameters, which depend on this fraction.  Relationships between the fraction of leaky junction $\phi$ and transport parameters are described thoroughly in \cite{our} and summarized below:
\begin{enumerate}
\item Endothelial effective diffusion coefficient $D_{end}^{eff}$:
\begin{equation}
    D_{end}^{eff}(\phi)=D_{lj}=D_{lumen}(1-\alpha_{lj})F(\alpha_{lj})\frac{4w}{R_{cell}}\phi,
    \label{eq:D_WSS}
\end{equation}
where $D_{lj}$ is the diffusion coefficient in the leaky junction and $D_{lumen} = 2.867 \times 10^{-11}
\frac{\mathrm{m^2}}{\mathrm{s}}$ is the diffusion coefficient in free plasma in the lumen. In our calculations we use the half width of a leaky junction $w$
equal to $14.3 \, \mbox{nm}$ (see Figure \ref{fig:leaky}). $\alpha_{lj}$ is the ratio of $r_{m}$ to $w$:
\begin{equation}
    \alpha_{lj}=\frac{r_{m}}{w},
\end{equation}
where $r_m$ is the radius of the LDL
molecule equal to $11 \, \mbox{nm}$, 
$(1-\alpha_{lj})$ is called the partition coefficient and $F$ is the hindrance factor,
\begin{equation}
F(\alpha_{lj})=1-1.004\alpha_{lj}+0.418\alpha_{lj}^{3}-0.169\alpha_{lj}^{5}.
\end{equation}

\item Endothelial permeability $K_{end}$: 
\begin{equation}
\label{eq:KK}
K_{end}(\phi)=K_{lj}(\phi)+K_{nj},
\end{equation}
where $K_{nj}$ is the normal junction permeability, which is independent on the fraction of leaky junctions
\begin{equation}
\label{eq:Knj}
K_{nj}=3.09\times 10^{-15} mm^2, 
\end{equation}
 
$K_{lj}$ is the permeability of leaky junctions dependent on the
fraction of leaky junctions $\phi$. It may be determined using formula
 
\begin{equation}
K_{lj}=\frac{w^{2}}{3}\frac{4w\phi}{R_{cell}}.
\end{equation}

\item Endothelial reflection coefficient $\sigma_{end}$ is calculated from the heteroporous model \cite{patlak1971theoretical, rippe1994transport, kellen2003transient}:
\begin{equation}
\label{eq:sigma}
\sigma_{end}=1-\frac{(1-\sigma_{lj})K_{lj}}{K_{end}},
\end{equation}
where $\sigma_{lj}$ is the  leaky junctions reflection coefficient given by
\begin{equation}
    \sigma_{lj}=1-(1-\frac{3}{2}\alpha_{lj}^{2}+\frac{1}{2}\alpha_{lj}^{3})(1-\frac{1}{3}\alpha_{lj}^{2}).
\end{equation}
\end{enumerate}

%------------------------------------------------
\subsection{Intima}

Intima is a cushion layer made of connective tissue, which may contain smooth muscle cells and few fibroblasts. This layer is relatively easily penetrable by both plasma and macromolecules \cite{vafai_yang_2006, our}. It should be noticed that  the abnormally high accumulation of LDL in the intima is associated with development of the atherosclerosis plague \cite{tarbell2003mass}.

The thickness of the intima layer in the four-layer models is assumed to be $10 \, \mu \mbox{m}$ \cite{vafai_yang_2006, prosi}. However, this value is different for various animals species and depends on the age of the specimen  \cite{atkinson1994association}. In this study we refer to the experimental profiles obtained by Curmi et al. \cite{curmi1990effect} and Meyer et al. \cite{meyer}. The intima of the rabbit aorta, which was the object of these experiments, is less than $5 \, \mu \mbox{m}$ thick. Thus, in our model we use $5 \, \mu \mbox{m}$ intima in case of physiological pressure $\Delta P=70 \, \mathrm{mmHg}$. In the hypertension case the intima thickness is reduced due to the compression effect.

The intima layer consists of proteoglycan and collagen components. Therefore, the heteroporous fiber matrix model, which takes into account both types of fibers, should be used to estimate intima transport properties, such as porosity, diffusion, reflection coefficient and permeability \cite{frank1989ultrastructure, chung2013arterial}. In the proteoglycan matrix, proteoglycan core proteins are connected with glycosaminoglycan fibers. They form a long central filament of hyaluronic.  The collagen is much thicker than any of proteoglycan components and its length per unit volume is much smaller. Therefore, the presence of the collagen is treated separately from that of the proteoglycan components \cite{dabagh2009transport, huang1994fiber}.

According to the fiber matrix model, the porosity of the intima can be calculated as
\begin{equation}
    \epsilon_{int}= \epsilon_{pg} \epsilon_{cg},
    \label{eq:eps_int}
\end{equation}
where $\epsilon_{pg}$ and $\epsilon_{cg}$ are porosities of the proteoglycan matrix and collagen fibers, respectively. For the physiological pressure we take $\epsilon_{pg} = 0.9568$ and $\epsilon_{cg} = 0.8387$ as in Dabagh et al. \cite{dabagh2009transport} and Liu et al. \cite{liu2011effect}.

The effective diffusivity of the intima can be obtained from the
equation \cite{dabagh2009transport}
\begin{equation}
    D_{int}^{eff}=D_{f} \, \exp\left[-(1-\epsilon_{pg})^{1/2}\left(1+\frac{r_m}{r*}\right)\right],
    \label{eq:D_int}
\end{equation}
where  $r_{m}$ is the radius of LDL molecule equal to $11 \, \mbox{nm}$, $r*$ is the effective radius for the entire proteoglycan matrix, $D_f$ is the diffusion coefficient of a solute in 
free space with compensation for the presence of the collagen,
\begin{equation}
    D_{f}=D_{lumen}(\epsilon_{cg}+\epsilon_{pg}-1)\exp\left[-(1-\epsilon_{cg})^{1/2}\left(1+\frac{r_m}{r_{cg}}\right)\right],
    \label{eq:D_f_int}
\end{equation}
where $r_{cg}$ is the radius of the proteoglycan fiber $(r_{cg}=20 \, \mbox{nm})$.

The effective radius for the entire proteoglycan matrix is given by the formula
\begin{equation}
    r*=\left[\frac{\alpha r_{mon}^2+r^2_{cf}}{\alpha+1}\right]^{1/2},
    \label{eq:reff_int}
\end{equation}
where $r_{cf}$ is the radius of central filament of hyaluronic acid and it equals $2 \, \mbox{nm}$. The effective monomer radius $r_{mon}$ of the proteoglycan core protein and glycosaminoglycans fibers is equal to
\begin{equation}
    r_{mon}=[ \beta r_{g}^{2}+r_{cp}^2 ]^{1/2},
    \label{eq:rm_int}
\end{equation}
where the proteoglycan core protein radius $r_{cp}$ is equal to $2 \, \mbox{nm}$,  $\alpha$ and $\beta$ are ratios of the total protein core to the central filament length and glycosaminoglycans to total protein core lengths, respectively. Typical values of $\alpha$ and $\beta$ should be in the range from $3$ to $10$ and $5$ to $10$, respectively. In the present study, we use $\alpha=3$, $\beta=5$, as in Dabagh et al. \cite{dabagh2009transport} and Liu et al. \cite{liu2011effect}.

The reflection coefficient of the intima is given by \cite{dabagh2009transport, liu2011effect}
\begin{equation}
    \sigma_{int}=(1-\Phi_f)^{2},
    \label{eq:sigma_int}
\end{equation}
where $\Phi_f$ is the partition coefficient
\begin{equation}
    \Phi_{f}=\exp\left[-(1-\epsilon_{pg})\left( 2\frac{r_{m}}{r*}+\frac{r_{m}^{2}}{r*^2}\right)\right](\epsilon_{cg}+\epsilon_{pg}-1)\exp\left[-(1-\epsilon_{cg})^{1/2}\left(1+\frac{r_{m}}{r_{cg}}\right)\right].
    \label{eq:phi_int}
\end{equation}

The hydraulic permeability of the intima has been determined as follows
\begin{equation}
\frac{1}{K_{int}}=\frac{1}{K_{pg}}+\frac{1}{K_{cg}}.
\label{eq:Kint}
\end{equation}
Particular permeabilities $K_{pg}$ and $K_{cg}$ can be calculated by utilizing the Carman-Kozeny equation \cite{curry1980fiber}
\begin{eqnarray}
K_{pg}=\frac{{r*}^2\epsilon_{pg}^{3}}{4G(\epsilon_{pg})(1-\epsilon_{pg})^{2}}, \\
\label{eq:K_pg}
K_{cg}=\frac{r_{cg}^{2}\epsilon_{cg}^{3}}{4G(\epsilon_{cg})(1-\epsilon_{cg})^{2}},
\label{eq:K_cg}
\end{eqnarray}
where $G$ is the Kozeny constant. For uncharged, randomly oriented cylindrical fibers it is given by \cite{huang, happel2012low,pluen1999diffusion}:
\begin{equation}
G(\epsilon)=\frac{2}{3}\frac{2 \epsilon^3}{(1-\epsilon)[2\ln(\frac{1}{1-\epsilon})-3+4(1-\epsilon)-(1-\epsilon)^2)]}+\frac{1}{3}\frac{2\epsilon^3}{(1-\epsilon)[\ln(\frac{1}{1-\epsilon})-\frac{1-(1-\epsilon)^2}{1+(1-\epsilon)^2})]}.
    \label{eq:G}
\end{equation}

%------------------------------------------------
\subsection{Internal Elastic Lamina}
Internal Elastic Lamina (IEL) seems to play a significant role in the LDL accumulation process. It is a selective permeable membrane constructed of impermeable connective tissue with
fenestral pores. This connective tissue is impermeable for both water and LDL. Thus, it form a significant barrier for LDL macromolecules \cite{dabagh2009transport, vafai_yang_2006}. Under certain conditions, IEL could be a stronger barrier to the LDL transport than endothelium \cite{our}.

The LDL accumulation process connected with IEL barrier is highly sensitive to transport parameters of this layer and therefore, they need to be carefully estimated. In this paper we calculate IEL transport parameters from the pore theory with cylindrical pores. We treat IEL as an impermeable layer with fenestral pores of radius $d=0.15 \, \mu \mbox{m}$, which are uniformly distributed over IEL \cite{dabagh2009transport, formaggia2010cardiovascular}. Fenestral pores are filled with the intima matrix \cite{huang, huang1994fiber, huang1997fiber}. The area fraction of fenestral pores, $f =3.49 \times 10^{-3}$, is taken from \cite{dabagh2009transport}.

The effective diffusion coefficient for IEL, which takes into account the fenestral pore structure, is given by \cite{michel1999microvascular}
\begin{equation}
    D_{IEL}^{eff}=D_{int}^{eff} \, f \left[2(1-\alpha_{IEL})^2-(1-\alpha_{IEL})^4\right]\left[1-2.1\alpha_{IEL}+2.09 \alpha_{IEL}^3-0.95 \alpha_{IEL}^5\right],
    \label{eq:D_IEL}
\end{equation}
where $\alpha_{IEL}=r_m/d$ is the ratio between the molecule radius and the radius of pores, and $D_{int}^{eff}$ is the diffusion coefficient in a medium filling fenestral pores, i.e. intima matrix. The reflection coefficient can be estimated in a similar way: 
\begin{equation}
    (1-\sigma_{IEL})=(1-\sigma_{int})(1-\sigma_{f}).
    \label{eq:sigma_IEL}
\end{equation}
According to \cite{michel1999microvascular}, the reflection coefficient $\sigma_{f}$ connected with fenestral pores structure is given by
\begin{equation}
    \sigma_{f}=\frac{16}{3}\alpha_{IEL}^2-\frac{20}{3}\alpha_{IEL}^3+\frac{7}{3}\alpha_{IEL}^4.
    \label{eq:sigma_f}
\end{equation}
The permeability of IEL is calculated as
\begin{equation}
    K_{IEL}=K_{int} f.
    \label{eq:K_IEL}
\end{equation}

%------------------------------------------------
\subsection{Media}
Media is the thickest layer considered in our model. It is made up of alternating layers of smooth muscle cells and elastic connective tissue. In this layer the absorption of LDL particles on the surface of smooth muscle cells takes place. This reaction is modeled by the first order decay of LDL, with the reaction rate coefficient $k$ \cite{vafai_yang_2006, our, prosi, morris}. This coefficient is different in various models \cite{olgac2008computational, ai_vafai_2006,vafai_yang_2006, prosi, morris} and in our calculations we use the value $k=3.197 \times 10^{-4}$ estimated by Prosi et al. in \cite{prosi}.

%------------------------------------------------
%------------------------------------------------
\section{Effect of hypertension}
\label{sec:pressure}

In the model presented here, LDL transport parameters of arterial wall layers depend on the pressure. We include three pressure induced mechanisms: increased water filtration through the wall, enhanced endothelium permeability due to the increased mitosis and apoptosis \cite{curmi1990effect,bratzler}, and mechanical compression of the intima \cite{huang}.

%------------------------------------------------
\subsection{Filtration velocity}
The filtration velocity can be calculated based on Equations (\ref{eq:velocity} - \ref{eq:pde2}). Nevertheless, these equations can be simplified. It should be noticed, that since the model is one-dimensional, for stationary state the filtration velocity is constant in the entire system. Because the increased pressure causes changes of tissue properties during the initial incubation time, the filtration is, generally speaking, time dependent. However, the adjustment of the filtration velocity is much faster process than the relevant timescale for the LDL transport. Therefore, in our model we assume that adjusting the velocity in the system is an instantaneous process and in each moment the filtration is given by Equation \ref{eq:pde1}. Then, using analogy to the Ohm's law for electrical circuit, pressure dependent filtration velocity $u$ can be expressed as:
\begin{equation}
\label{eq:filtr}
	 u  = \frac{\Delta P}{\sum_{i=1}^4 \frac{L_i \mu}{K_i} },
\end{equation}
where $\Delta P$ is the transmural pressure, $L_i$ is the i-th layer thickness, $\mu$ is the fluid dynamic viscosity and $K_i$ is the i-th layer permeability.

%------------------------------------------------
\subsection{Changes in the endothelium}
In \cite{wu1990transendothelial} Wu et al. found that mitosis and apoptosis of endothelial cells increase under hypertension. Also Bretherton et al. \cite{bretherton1976effect} suggested that hypertension in the normally fed rabbit increases lipoproteins entry into the arterial wall because of the permeability rather than by a direct effect of filtration process. 

Upon these observations, the fraction of leaky junctions, which in our previous model \cite{our} was dependent only on the wall shear stress, is supplemented by the pressure and time dependent term
$\phi_P$
\begin{eqnarray}
   \label{eq:phi0}
      \phi\left( WSS, \Delta P\left[\mathrm{mmHg}\right], t\right)=\phi_{WSS}\left(WSS\right)+\phi_P\left(\Delta P\left[\mathrm{\mathrm{mmHg}}\right], t\right).
\end{eqnarray} 
Here, we assume that for physiological pressure $\phi_P(70 \, \mathrm{mmHg})=0$. 

The quantitative relationship between the pressure and the fraction of leaky junctions is not described in the literature. In our model we estimate values of $\phi_P$ for two levels of elevated pressure and the curve of the fraction of leaky junctions versus time of exposition to increased pressure $\phi(t)$. In order to obtain it, we fit our results to experimentally obtained averaged tissue concentration profiles. We use the least square method with the $\phi_P$ as a free parameter. 

%------------------------------------------------
\subsection{Intima compression}
The other hypertension effect is the intima mechanical compression discussed by Huang et al. in \cite{huang}. Olgac et al. in \cite{olgac2009patient} pointed out that this effect should be taken into account in calculations of the LDL transport in the arterial wall in the case of elevated pressure. This effect was considered by Dabagh et al. in \cite{dabagh2009transport} and introduced to the four-layer model by Liu et al. in \cite{liu2011effect}. 

Increased transmural pressure compresses the intima and hence changes porosities of the proteoglycan matrix $\epsilon_{pg}$ and collagen fibers $\epsilon_{cg}$. These porosities are functions of pressure dependent intima thickness
\begin{eqnarray}
\epsilon_{pg}=1-\frac{1-\epsilon_{0pg}}{cf}, \\
\label{eq:epsilon_pg}
\epsilon_{cg}=1-\frac{1-\epsilon_{0cg}}{cf},
\label{eq:epsilon_cg}
\end{eqnarray}
where $\epsilon_{0pg}$ and $\epsilon_{0cg}$ are porosities at zero transmural pressure, equal to $0.9866$ and $0.95$, respectively \cite{dabagh2009transport}. The compression factor $cf$ is defined as the ratio of the intima thickness at zero luminal pressure $L_0$ to its thickness at  given pressure $L_i$: $cf=L_i/L_0$. Values of $cf$ for different pressure levels are adopted from \cite{huang1997fiber}. Changes in porosities influence all transport parameters of the intima as described in details in the previous section.

%------------------------------------------------
%------------------------------------------------
\section{Experiments}
In order to estimate the pressure and time dependent fraction of leaky junctions $\phi(\Delta P, t)$ we use experimental results provided by Curmi et al. \cite{curmi1990effect} and Meyer et al. \cite{meyer}. They performed experiments for LDL transport in the arterial wall under normal and hypertension cases with transmural pressures equal to $70 \, \mathrm{mmHg}$, $120 \, \mathrm{mmHg}$ (only Meyer et al.) and $160 \, \mathrm{mmHg}$. Moreover, Curmi et al. investigated the LDL transport under transmural pressure equal to $160 \, \mathrm{mmHg}$ with several times of exposition to elevated pressure.

Both mentioned experiments were very similar. The unique procedure of in vitro preparation allowed to perform experiments with intact rabbit thoracic aorta. Separated arteries were incubated with the intraluminal solution containing \ce{^{131}I-LDL} molecules. The transmural distribution of relative concentrations of LDL across the wall was determined by using a serial frozen sectioning technique. En face $20\mbox{-}\mu \mbox{m-}$thick serial sections were cut through the whole thickness of the wall, from the endothelium to the adventitia. It should be mentioned that the first $20\mbox{-}\mu \mbox{m-}$thick section contained endothelium, intima, IEL and part of the inner media. For each section the radioactive \ce{^{131}I} was estimated with a double counting procedure on a gamma counter. Relative LDL concentration was calculated as counts per minute per unit volume of wet tissue, divided by counts per minute per unit volume of intraluminal solution. Obtained relative LDL tissue concentrations form a profile, which can be directly compared with appropriate model results.
 
To summarize, to obtain the fraction of leaky junctions we use the experiments performed under following conditions: 
\begin{itemize}
    \item under $70 \, \mathrm{mmHg}$, $120 \, \mathrm{mmHg}$ and $160 \, \mathrm{mmHg}$ with $30$ minutes incubation, performed by Meyer et al.,
    \item under $160 \, \mathrm{mmHg}$ with incubation lasting for $15$ minutes, $30$ minutes, $1$ and $2$ hours, performed by Curmi et al. 
\end{itemize}    
    Curmi et al. showed that results obtained already after 1 hour can be treated as a stationary state.

%------------------------------------------------
%------------------------------------------------
\section{Mathematical model and parameters}
In order to obtain the LDL concentration profile we need to solve the system of three Equations (\ref{eq:velocity} - \ref{eq:pde2}). Similarly as in \cite{our}, we solve transport equations in one dimension. Therefore, for each layer we solve numerically the one-dimensional PDE:
\begin{equation}
\label{eq:transport1d}
      \epsilon_i\frac{\partial c}{\partial t}    =   D^{eff}_i\frac{\partial^2 c}{\partial x^2} -(1-\sigma_i) u \cdot \frac{\partial c}{\partial x} - k_i c
\end{equation}
At interfaces between layers the flux continuity condition is applied
\begin{equation}
\left.\left[(1-\sigma_i){u}c-D^{eff}_i\frac{\partial c}{\partial x}\right]\right|_{+}=\left.\left[(1-\sigma_j){u}c-D^{eff}_j\frac{\partial c}{\partial x}\right]\right|_{-},
     \label{eq:continuity}
\end{equation}
where + and - denote particles flux at the left and right side of the boundary, respectively. The filtration velocity is calculated from Equation (\ref{eq:filtr}).

Additionally, Dirichlet boundary conditions for $c(0)$ and $c(l)$ were used in computations ($l$ is the total length of calculation domain). The value of the left boundary $c(0)$ is taken to be $1$. Let us recall that due to the linearity of the transport equation we can use the relative concentration, where $c=1$ is in the lumen \cite{our}. The right  boundary condition $c(l)$ is taken from experimental LDL concentration at the interface between the media and adventitia \cite{meyer, curmi1990effect}. This value is different for various transmural pressures. However, it has to be stressed that the $c(l)$ value has very small effect on the concentration profile, since the most of LDL molecules is absorbed in the media layer \cite{vafai_yang_2006}.

Layers of the arterial wall are characterized by several parameters, i.a. layer thickness, effective diffusivity, reflection coefficient, permeability, viscosity and porosity. Parameters of endothelium and intima depend on transmural pressure. Moreover, parameters of endothelium depend also on time. Stationary state parameters used in this paper are summarized in Table \ref{tab:parameters}.

The details of the model implementation
as well as the full source code can be found on Git repository \cite{gitLDL}.

% TABLE WITH PARAMETERS
\begin{table}[]
\centering
\begin{tabular}{|l|r|c|c|c|c|}

 \hline
 \multicolumn{2}{|c|}{\textbf{Parameter}} & \textbf{Endothelium} & \textbf{Intima} & \textbf{IEL} & \textbf{Media} \\
 \specialrule{1pt}{0pt}{0pt}%\hline
 % 
 % ^^^^^^^^^^^
  \multirow{3}{*}{{\specialcell{\textbf{Thickness} \\ \textit{L, $\mu m$}}}}&\multirow{1}{*}{{\scriptsize{$70 \mathrm{mmHg}$}}} & \multirow{3}{*}{{$2.0$}} & \multirow{1}{*}{{$5.0$}} & \multirow{3}{*}{{$2.0$}} & \multirow{3}{*}{{$161.0$}} \\
  \cline{2-2} 
  \cline{4-4}
   & \multirow{1}{*}{{\scriptsize{$120 \mathrm{mmHg}$}}} & & \multirow{1}{*}{{$2.3$}} & &  \\   
   \cline{2-2} 
  \cline{4-4}
   & \multirow{1}{*}{{\scriptsize{$160 \mathrm{mmHg}$}}} & & \multirow{1}{*}{{$2.0$}} & &  \\ \specialrule{1pt}{0pt}{0pt}%\hline
 \multirow{3}{*}{{\specialcell{\textbf{Diffusivity}\\ \textit{$D^{eff}$, $mm ^{2}/s$}}}} & \multirow{1}{*}{{\scriptsize{$70 \mathrm{mmHg}$}}} & $4.7 \times 10^{-12}$ & $3.68 \times 10^{-6}$ & \multirow{3}{*}{$1.066 \times 10^{-8}$} & \multirow{3}{*}{$5.0 \times 10^{-8}$} \\ \cline{2-4}
  &  \multirow{1}{*}{{\scriptsize{$120 \mathrm{mmHg}$}}} & $ 5.82\times 10^{-11}$ & $1.08 \times 10^{-6}$ & & \\ \cline{2-4}
  &  \multirow{1}{*}{{\scriptsize{$160 \mathrm{mmHg}$}}} & $ 1.03 \times 10^{-10}$ & $8.263 \times 10^{-7}$ & & \\ \specialrule{1pt}{0pt}{0pt}%\hline
  \multirow{3}{*}{{\specialcell{\textbf{Reflection} \\\textbf{coefficient} \textit{$\sigma$}}}}& \multirow{1}{*}{{\scriptsize{$70 \mathrm{mmHg}$}}} & $0.9890$ & $0.7998$ & \multirow{3}{*}{$0.8051$} & \multirow{3}{*}{$0.8836$} \\ \cline{2-4}
  &  \multirow{1}{*}{{\scriptsize{$120 \mathrm{mmHg}$}}} & $ 0.9064$ & $0.9789$ & & \\ \cline{2-4}
  &  \multirow{1}{*}{{\scriptsize{$160 \mathrm{mmHg}$}}} & $ 0.8688$ & $0.9876$ & & \\ \specialrule{1pt}{0pt}{0pt}%\hline
  \multirow{3}{*}{{\specialcell{\textbf{Permeability}\\\textit{K, $mm^{2}$} }}} & \multirow{1}{*}{{\scriptsize{$70 \mathrm{mmHg}$}}} & $3.22 \times 10^{-15}$ & $3.907 \times 10^{-11}$ & \multirow{3}{*}{$1.36 \times 10^{-13}$} & \multirow{3}{*}{$2.0 \times 10^{-12}$} \\ \cline{2-4}
  &  \multirow{1}{*}{{\scriptsize{$120 \mathrm{mmHg}$}}} & $4.7 \times 10^{-15}$ & $9.91\times 10^{-12}$ & & \\ \cline{2-4}
  &  \multirow{1}{*}{{\scriptsize{$160 \mathrm{mmHg}$}}} & $5.95 \times 10^{-15}$ & $7.72\times 10^{-12}$ & & \\ \specialrule{1pt}{0pt}{0pt}%\hline
  \multirow{3}{*}{\textbf{Porosity} \textit{$\epsilon$}} & \multirow{1}{*}{{\scriptsize{$70 \mathrm{mmHg}$}}} & $1.91\times 10^{-6}$ & $0.8025$ & \multirow{3}{*}{$0.003$} & \multirow{3}{*}{$0.258$} \\ \cline{2-4}
  & \multirow{1}{*}{{\scriptsize{$120 \mathrm{mmHg}$}}} & $2.36\times 10^{-5}$ & $0.5862$ & & \\ \cline{2-4}
  & \multirow{1}{*}{{\scriptsize{$160 \mathrm{mmHg}$}}} & $4.19\times 10^{-5}$ & $0.5401$ & & \\ \specialrule{1pt}{0pt}{0pt}%\hline
  \multicolumn{2}{|l|}{\textbf{Viscosity}} &  \multicolumn{4}{c|}{\multirow{2}{*}{$0.72 \times 10^{-3}$ }}\\ 
  \multicolumn{2}{|l|}{\textit{$\mu$, $g/(mm\cdot s)$}}&  \multicolumn{4}{c|}{} \\ \specialrule{1pt}{0pt}{0pt}%\hline

\end{tabular}

\caption{Stationary state parameters of the arterial wall used in calculations.}
\label{tab:parameters}
\end{table}

% END OF PARAMETERS TABLE

%------------------------------------------------
%------------------------------------------------
\section{Results} \label{results}

The key parameter for the LDL transport studied in this paper is the transmural pressure. As we have already mentioned, calculations for three values of transmural pressure:  $70 \, \mathrm{mmHg}$, $120 \, \mathrm{mmHg}$  and $160 \, \mathrm{mmHg}$ have been performed. It corresponds to values in experiments conducted by Meyer et al. \cite{meyer} and Curmi et al. \cite{curmi1990effect}. Those experiments were performed under physiological WSS, therefore we use $WSS = 2.5$ in all our simulations. This value of wall shear stress corresponds to the fraction of leaky junctions $\phi_{WSS} = 5.0 \times 10^{-4}$, as in \cite{yiannis2007role, dabagh2009transport}. 

Since we use experimental results to estimate the fraction of leaky junctions, we have to pay special attention to compare appropriate quantities with each other. Experiments and theoretical simulations result in different outcomes. Computational calculations provide information about the LDL concentration profile in the fluid (plasma). In contrast, experiments give the LDL tissue concentration averaged over $20\mbox{-}\mu\mbox{m-}$slices \cite{meyer, curmi1990effect}. Therefore, we converted the theoretical fluid concentration into the averaged tissue concentration. In the first step we transform the fluid volume into the tissue volume by multiplying it by the appropriate porosity of the layer. In the second step the tissue concentration is averaged over $20\mbox{-}\mu \mbox{m-}$slices. Results of each transformation step can be seen in Figure \ref{fig:70}.  

This section is divided into four parts. In the first three subsections we present separately results for each analyzed pressure level. In the last part, LDL concentrations for all values of pressure are summarized and the impact of each pressure induced effect is discussed. In all figures presented in this paper concentration profiles in the fluid are marked by solid lines, concentration profiles in the tissue by dashed lines, averaged tissue concentrations by dots or triangles, and experimental data by crosses, unless otherwise noted. We also follow the convention that a marker denotes an average concentration over a $20\mbox{-}\mu \mbox{m-}$slice on the left from its $x$-position. Curmi additionally corrected the position of measurements in order to account for the real slice width, which can be noticed in plots as deviation from multiplicities of $20 \, \mu m$ \cite{curmi1990effect}.  

%------------------------------------------------
\subsection{LDL concentration under physiological conditions - $70 \, \mathrm{mmHg}$}
\label{sec:70}

\begin{figure}[t] 
\centering
\includegraphics[width=1.0\textwidth, keepaspectratio=true]{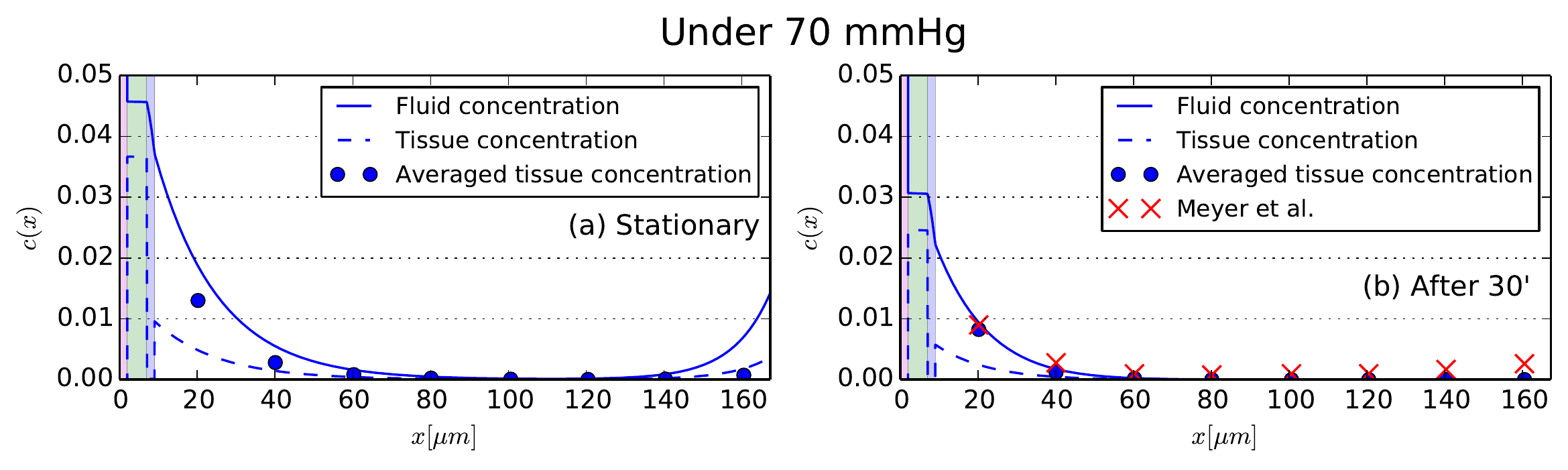}
\caption{LDL concentration profiles across the arterial wall obtained for transmural pressure equal to  $70 \, \mathrm{mmHg}$: (a) stationary state and (b) after 30 minutes of incubation. '\textit{Fluid concentration}'(\textcolor{Blue}{$\boldsymbol{-}$})
is the normalized LDL concentration in the plasma, obtained directly from calculations. '\textit{Tissue concentration}'  (\textcolor{Blue}{\textbf{\texttt{-{}-}}})
is the normalized LDL concentration profile in the wet tissue. '\textit{Averaged tissue concentration}' ($\filledcirc{blue}$)
is the tissue concentration averaged in the $20\mbox{-}\mu \mbox{m-}$slices as in experiments. Results after 30 minutes (b) are put together with experimental results of Meyer et al. \cite{meyer} ({\color{red}$\boldsymbol{\times}$}). Subsequent layers are marked with colors: magenta for endothelium, green for intima, blue for IEL and white for media.   
}
\label{fig:70}
\end{figure}

The physiological condition corresponds to the transmural pressure equal to  $70 \, \mathrm{mmHg}$. In this case we can compare our theoretical LDL concentration profiles with experimental results provided by Meyer et al. \cite{meyer}. As mentioned before, the experiment was performed only for $30$ minutes. Unfortunately, for $70 \, \mathrm{mmHg}$ there is no experimental data for stationary state. Therefore, we compared our theoretical LDL concentration profile with results obtained by Meyer et al. for $30$ minutes incubation. Moreover, we theoretically calculated the LDL concentration profile for the stationary state. Results  are presented in Figure \ref{fig:70}. It can be noticed that stationary state is not reached after only $30$ minutes. The very good agreement between Meyer's data and non-stationary LDL concentration profile can be noticed. 

%------------------------------------------------
\subsection{LDL concentration in the arterial wall under $160 \, \mathrm{mmHg}$}
For the transmural pressure equal to $160 \, \mathrm{mmHg}$ effects connected with the hypertension become significant. Moreover, in this case there are several experiments which provide valuable data about LDL profiles for different times of incubation.
%We can use it in our considerations.

\begin{figure}[t] 
\centering
\includegraphics [width=1.0\textwidth, keepaspectratio=true]{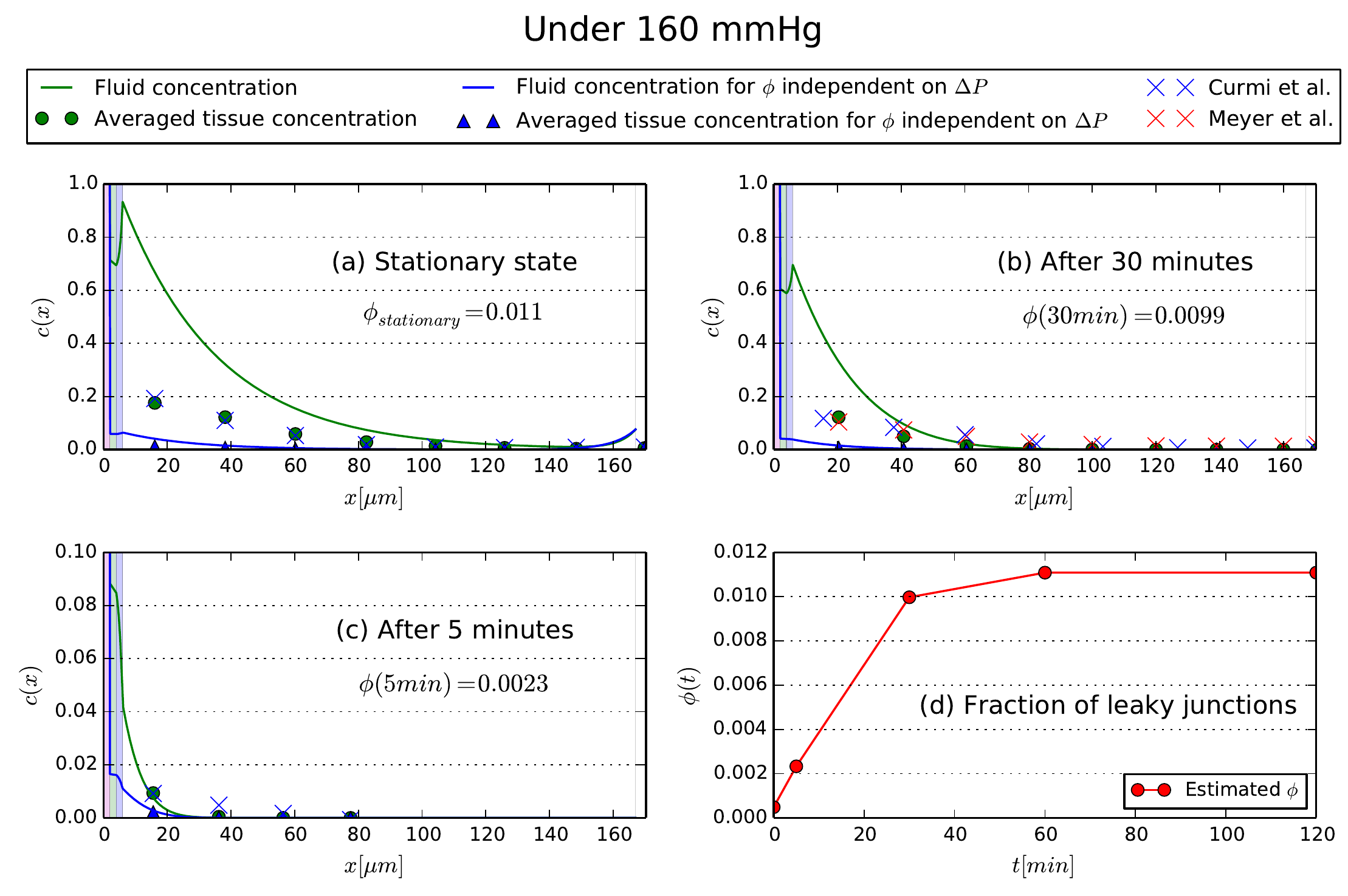}
\caption{ LDL concentration profiles across the arterial wall obtained for transmural pressure equal to  $160 \, \mathrm{mmHg}$: (a) stationary state, (b) after 30 minutes of incubation, (c) after 5 minutes of incubation.  Fluid concentration with fraction of leaky junction dependent on pressure and time is marked by (\textcolor{OliveGreen}{$\boldsymbol{-}$}), averaged tissue concentration with $\phi$ dependent on pressure and time is marked by ($\filledcirc{black!50!green}$), fluid concentration with fraction of leaky junction independent on pressure and time is marked by (\textcolor{Blue}{$\boldsymbol{-}$}) and averaged tissue concentration with $\phi$ independent on pressure and time is marked by ($\textcolor{Blue}{\blacktriangle}$). Results are put together with experimental results of Meyer et al. \cite{meyer} ({\color{red}$\boldsymbol{\times}$}) and Curmi et al \cite{curmi1990effect} ({\color{blue}$\boldsymbol{\times}$}). Subsequent layers are marked with colors: magenta for endothelium, green for intima, blue for IEL and white for media. Time dependence of fraction of leaky junctions $\phi$ ($\filledcirc{red}$) is illustrated by the red curve (\textcolor{red}{---}) in Panel (d).}
\label{fig:160}
\end{figure}

As we already mentioned, there is no reasonable information about the fraction of leaky junctions $\phi$ under hypertension. Therefore, first we checked the outcome of calculations with the same value of $\phi$ as under physiological conditions, effectively treating $\phi$ as being independent on the transmural pressure. Results are denoted by blue colour in  Figure \ref{fig:160}. It can be seen that in this case theoretical results strongly underestimate experimental values. However, in the case of 5 minutes of incubation this underestimation is smaller than for 30 minutes and for stationary state. This fact suggests that the fraction of leaky junctions $\phi$ in the case of hypertension may depend not only on the pressure, but also on time of the exposure to high pressure, i.e. duration of incubation. This confirms Dabagh et al. suggestions from \cite{dabagh2009transport}.

On the base of this conclusion, we need to find the time dependence of the fraction of leaky junctions. In order to keep our model as simple as possible, we apply following iterative procedure. The first assumption is that the $\phi(t)$ is a piecewise linear function of time on intervals $(t_j,t_{j+1})$, where  $t_j\in $ \{$0$ min, $5$ min, $30$ min, $60$ min, $120$ min\} is experimentally investigated incubation time. For incubation time equal to $0$ we assume physiological value $\phi(0 \, \mathrm{min})=5\times 10^{-4}$. On the other hand, from Curmi experiments we know that after one hour the system is already in the stationary state. Thus, the $\phi(60 \, \mathrm{min})$ and $\phi(120 \, \mathrm{min})$  should be the same. We obtain them by finding the optimal value of $\phi$ for which the numerically calculated concentration profile matches the experimental data. We use the least square method and solve the stationary transport equation in each iteration step. The remaining two values $\phi(5 \, \mathrm{min})$ and $\phi(30 \, \mathrm{min})$ are estimated by solving the time dependent transport equation with time dependent coefficients. First, we solve it for $t\in(0\, \mathrm{min}, 5 \, \mathrm{min})$ and find an optimal linear dependence on this interval. Having fixed value of $\phi(5 \, \mathrm{min})$,  we obtain $\phi(30 \, \mathrm{min})$ in the same way.  As a result we determine the time dependence $\phi(t)$, which is presented in Figure \ref{fig:160}d. 

Concentration profiles obtained numerically with time dependent $\phi(t)$ and experimental results are presented in Figure \ref{fig:160}a-c. The very good agreement validates our approach. Let us note, that the rate of endothelial cells apoptosis and mitosis adapts to the hypertension conditions in first 60 minutes of the exposure. Interestingly, in the modelling the process of LDL transport in the arterial wall for time independent $\phi$ (at $160 \, \mathrm{mmHg}$) reaches the stationary state after about 30 minutes, i.e. faster then it was observed in experiments. Therefore, we conclude that temporal changes in the tissue structure significantly delay the stabilization of the LDL concentration to about 1 hour.

%------------------------------------------------
\subsection{LDL concentration in the arterial wall under $120 \, \mathrm{mmHg}$}

\begin{figure}[t] 
\centering
\includegraphics [width=1.0\textwidth, keepaspectratio=true]{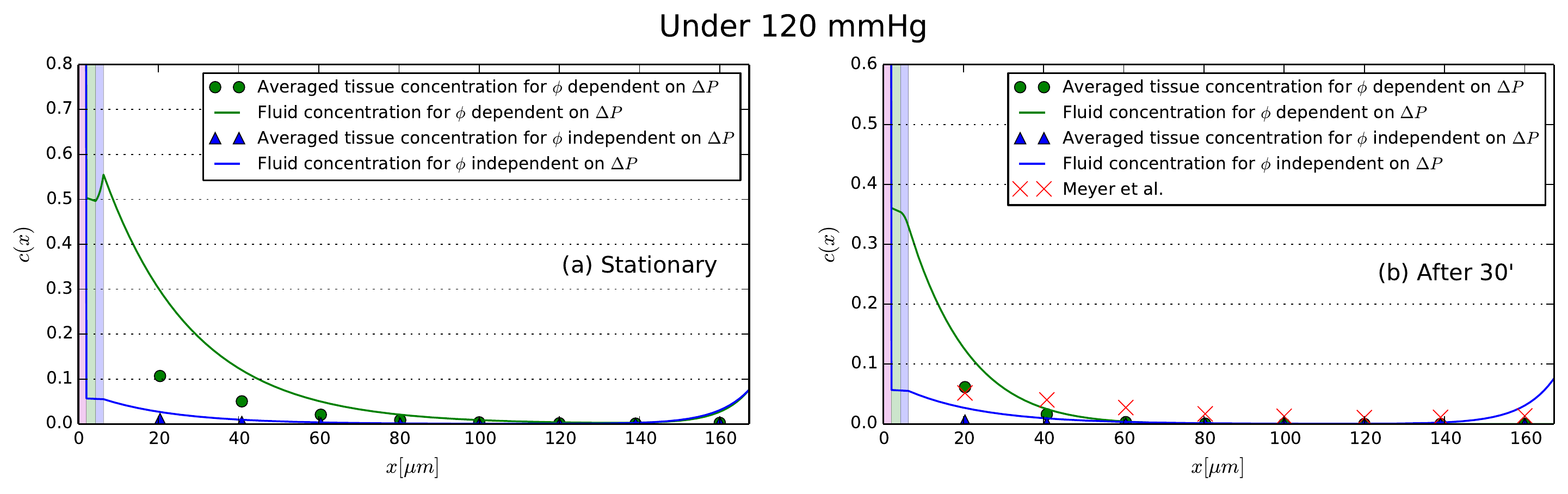}
\caption{Summary of LDL concentration profiles across the arterial wall obtained for transmural pressure equal to  $120 \, \mathrm{mmHg}$: (a) stationary state and (b) after 30 minutes of incubation. The  fluid concentration and averaged tissue concentration obtained with pressure dependent fraction of leaky junctions (\textcolor{OliveGreen}{---} $\filledcirc{black!50!green}$)  and  without pressure dependent fraction of leaky junctions  ({\color{blue}--- $\blacktriangle$}) are put together with the experimental results of Meyer et al. \cite{meyer}  ({\color{red}$\boldsymbol{\times}$}). The  subsequent layers are marked with colors: magenta for endothelium, green for intima, blue for IEL and white for media.}
\label{fig:120}
\end{figure}

Similarly to the case of physiological pressure, for the transmural pressure equal to $120 \, \mathrm{mmHg}$ we have experimental data obtained only after 30 minutes of incubation. Again, if we assume that the fraction of leaky junctions $\phi$ is independent on the pressure, the theoretical model is not able to reproduce results obtained experimentally, what can be seen in Figure \ref{fig:120}. In this case we assume that qualitatively the dependence of the fraction of leaky junctions on the incubation time is the same as for the case with transmural pressure equal to $160 \, \mathrm{mmHg}$. Thus, we scale the function $\phi(t)$ under  $160 \, \mathrm{mmHg}$ from Figure \ref{fig:160}d. The scaling factor is estimated by fitting theoretical data with 30 minutes incubation time to the experimental data provided by Meyer et al. \cite{meyer}. Our model predictions for stationary state and for 30 minutes incubation are shown in Figure \ref{fig:120}. In this case we can compare numerical model with experiment only for the transient concentration profile. The agreement is not as good as under $160 \, \mathrm{mmHg}$ case, nevertheless the model predicts correctly the order of magnitude. 

%------------------------------------------------
\subsection{Effect of the hypertension on the LDL concentration in the arterial wall}
\begin{figure}[t] 
\centering
\includegraphics [width=1.0\textwidth, keepaspectratio=true]{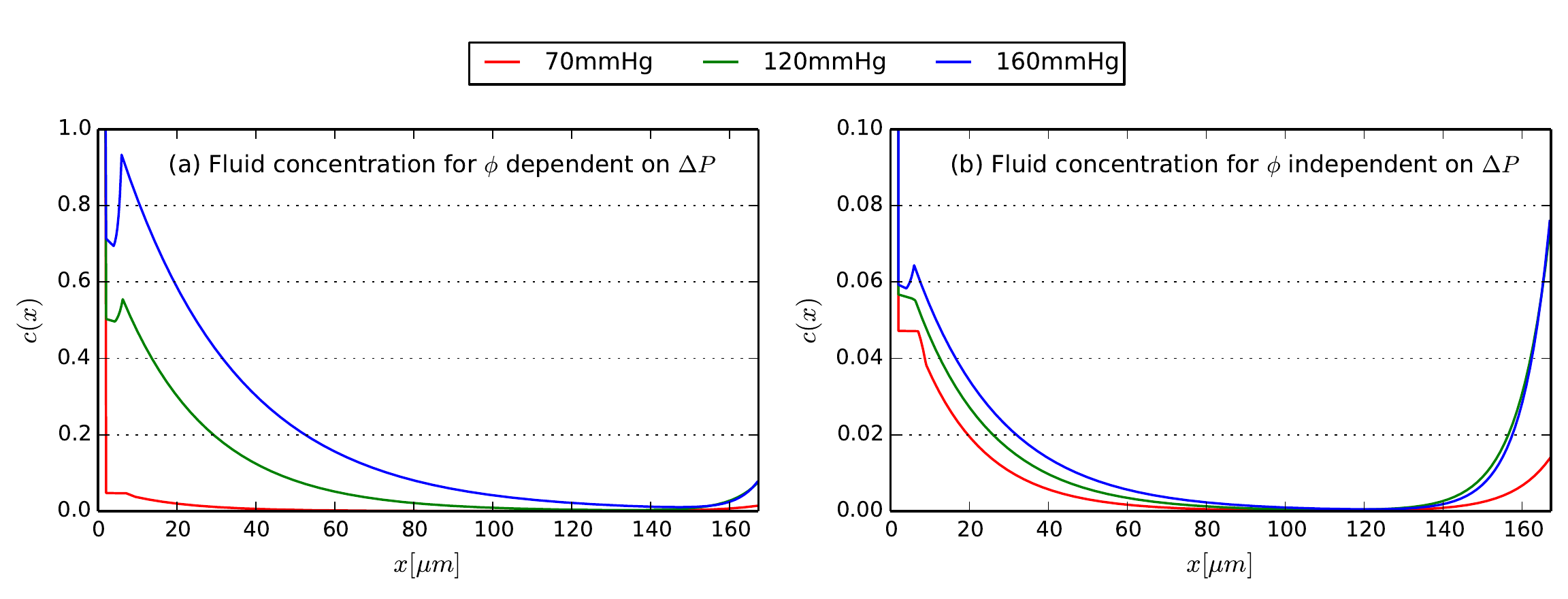}
\caption{Fluid concentration profiles across the arterial wall in the stationary state under different pressure conditions: $\Delta P = 70 \, \mathrm{mmHg}$ (\textcolor{red}{$\boldsymbol{-}$}), $\Delta P = 120 \, \mathrm{mmHg}$ (\textcolor{OliveGreen}{$\boldsymbol{-}$}), $\Delta P = 160 \, \mathrm{mmHg}$ (\textcolor{Blue}{$\boldsymbol{-}$}): (a) fraction of leaky junction $\phi$ dependent on the $\Delta P$ and (b) fraction of leaky junction $\phi$ independent on the $\Delta P$.}
\label{fig:pressure}
\end{figure}

Experiments performed by Curmi et al. \cite{curmi1990effect} and Meyer et al. \cite{meyer} predict significant impact of the pressure on  LDL concentration profiles. In Figure \ref{fig:pressure} we summarized obtained theoretical LDL concentration profiles with fraction of leaky junctions dependent on the pressure (Figure \ref{fig:pressure}a) and without this effect (Figure \ref{fig:pressure}b). It can be seen that it is crucial to take into account pressure dependent fraction of leaky junctions. Neglecting this effect results in underestimation of the LDL concentration by one order of magnitude. 
Effective values of the fraction of leaky junctions in stationary state, estimated for each value of the transmural pressure are shown in Table \ref{tab:relative160}. It can be noticed that the fraction of leaky junction is directly proportional to the transmural pressure. Let us remind that dependence of the fraction of leaky junctions on the pressure level under hypertension were not so far described in the literature and it is usually assumed to be in the wide range between $0.001$ and $0.04$. Our predicted values are within this range. 
\begin{table}[ht]
\centering
\begin{tabular}{|l|c|}
\hline
Pressure $\Delta P$ &  Fraction of leaky junctions $\phi$ \\ \hline
$70 \, \mathrm{mmHg}$      &   $5\times 10^{-4} $                     \\ \hline
$120 \, \mathrm{mmHg}$                               &  $6.2\times 10^{-3} $                    \\ \hline
$160 \, \mathrm{mmHg}$                 & $1.1\times 10^{-2} $                    \\ \hline
\end{tabular}
\caption{Predicted fraction of leaky junctions $\phi$ for different values of pressure (stationary states).  }
\label{tab:relative160}
\end{table}

\begin{figure}[t] 
\centering
\includegraphics [width=0.5\textwidth, keepaspectratio=true]{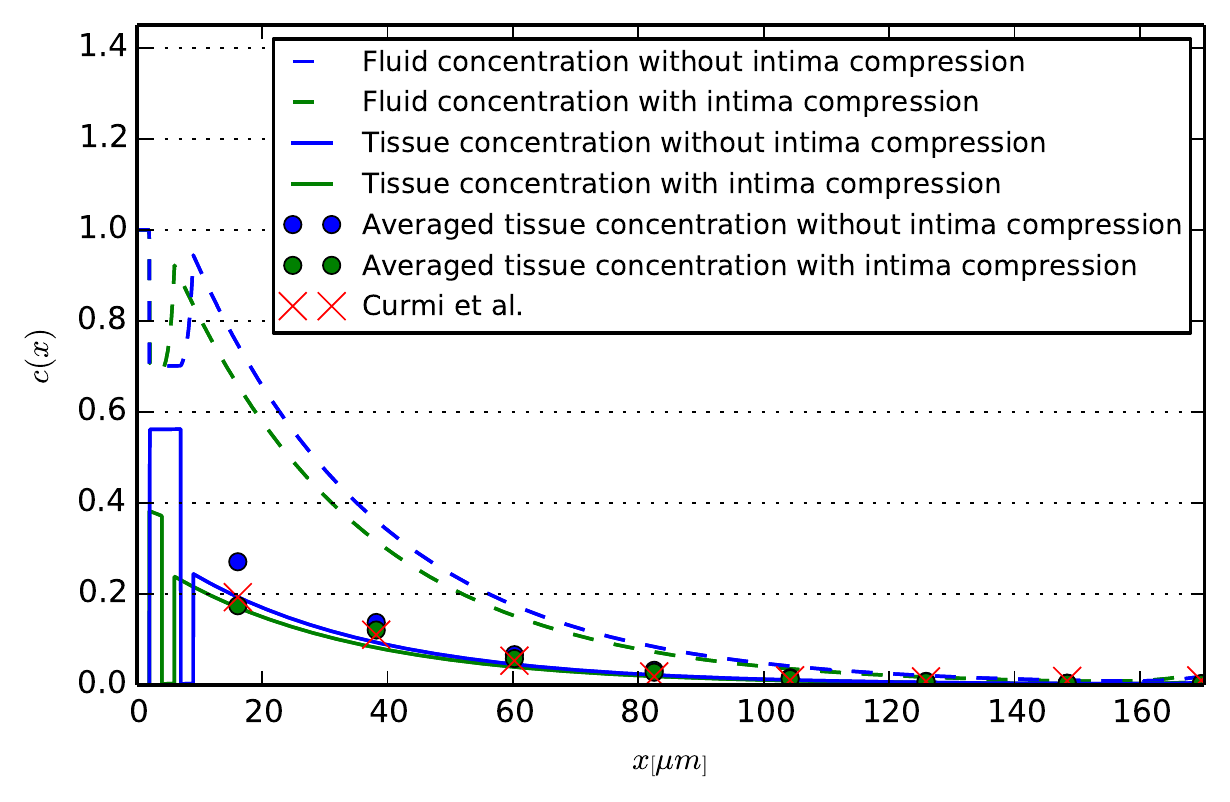}
\caption{Profiles of normalized LDL concentration across the arterial wall obtained with included intima compression (averaged tissue concentration ($\filledcirc{black!50!green}$), fluid
concentration (\textcolor{OliveGreen}{ $\boldsymbol{-}$}) and tissue concentration (\textcolor{OliveGreen}{ \textbf{\texttt{-{}-}}})), and without intima compression
(averaged tissue concentration ($\filledcirc{blue}$), fluid concentration ({\color{blue} $\boldsymbol{-}$}) and tissue concentration (\textcolor{blue}{ \textbf{\texttt{-{}-}}})), compared with experimental results of Curmi et al. \cite{curmi1990effect} ({\color{red}$\boldsymbol{\times}$}). }
\label{fig:160compression}
\end{figure}

In contrast to the effect of the pressure dependent fraction of leaky junctions, the role of intima compression is not so dramatic. However, if we neglect this effect, the agreement with the experiments becomes visibly worse.  For transmural pressure equal to $160 \, \mathrm{mmHg}$ the experimental averaged tissue concentration profile  become slightly flattened in comparison to  $70 \, \mathrm{mmHg}$. It could be connected with the intima compression effect, what seems to be confirmed by results obtained with our model presented in Figure \ref{fig:160compression}. It can be noticed that mechanical compression of the intima causes flattening of the averaged tissue concentration profile. In Figure \ref{fig:160compression} we present LDL concentration profiles calculated with and without compressed intima. In case of uncompressed intima, the parameters are the same as for the transmural pressure equal to $70 \, \mathrm{mmHg}$  (Table \ref{tab:parameters}). The profile with compressed intima is exactly the same as in Figure \ref{fig:160}a.

It can be seen that effect of the intima compression on averaged tissue concentration is mostly indirect. Namely, in individual layers the LDL concentration in fluid is on similar level in both cases. However, since the compression of the intima decreases its porosity, one could expect the decrease of the tissue concentration. Indeed, in Figure \ref{fig:160compression} one can notice this effect.  
 
%------------------------------------------------
%------------------------------------------------ 
\section{Conclusions}

In this paper we studied the influence of the hypertension on the LDL transport in the layered arterial wall. The problem of the LDL transport in the presence of the hypertension is a complex issue. We propose to use a four-layer model of the LDL transport supplemented by hypertension effects. The most of parameters of our model have been calculated based on the physiological structure of the arterial wall. Remaining parameters were taken from the literature \cite{vafai2012, prosi}. The question was whether this theoretical model can predict LDL concentration profiles reported in experiments for various levels of the transmural pressure. We also wanted to find out effects, which can not be neglected in the hypertension case. 

Pressure induced effects reported in the literature are: increased filtration velocity, mechanical intima compression and increased number of leaky junctions. All those effects were covered in our model. Conditions of our simulations have been chosen to correspond as closely as possible to experimental ones. Therefore, we have taken into account both stationary as well as time dependent LDL transport process under physiological pressure and in the case of hypertension. 

In the first step we test our model on the base of the physiological pressure. The very good agreement between our model end experimental data confirms its validity.

Then we performed calculation for the transmural pressure equal to $160 \, \mathrm{mmHg}$ and $120 \, \mathrm{mmHg}$. If the only effect of increased pressure would be the increased water filtration through the wall, then the discrepancy of LDL concentration level between models and experiments reaches one order of magnitude. Therefore, in the second step we took into account two other mentioned effects connected with the hypertension. 

The first and dominant one is the change in the endothelium structure. The elevated pressure causes increase in the fraction of leaky junctions $\phi$ and hence a higher permeability of the first layer, as well as increase in macromolecular influx into the wall \cite{wu1990transendothelial, bretherton1976effect}. Moreover,  the fraction of leaky junctions seems to be dependent on the time of exposure to the hypertension.
Since there is no valid data about values of the fraction of leaky junctions under hypertension, we estimated the time and pressure dependence of the fraction of leaky junctions. This estimation allows to reproduce the experimental LDL concentration profiles for all considered times of incubation, for both $160 \, \mathrm{mmHg}$ and $120 \, \mathrm{mmHg}$. This is the major finding in our paper: neglecting pressure dependent changes of fraction of leaky junctions causes significant underestimation of the LDL concentration in the arterial wall. This conclusion confirms the suggestions of the Bretherton et al. from \cite{bretherton1976effect} that hypertension in the normal fed rabbit increases lipoprotein entry into the arterial wall by an effect of vessel wall permeability rather than by a direct effect of filtration velocity.  Moreover, we  determined the time dependence of the fraction of leaky junctions. The time of vessel reaction is longer than 30 minutes.  It means that the impact of the hypertension on the arterial wall is indirect and is connected with the increase in the mitosis and apoptosis of the endothelial cells. To the best of authors' knowledge, the dependence of the fraction of leaky junctions on the pressure level and on the time of exposition to the hypertension have not been quantitatively described in the literature.

The second effect which have been taken into account is the compression of intima layer caused by hypertension. This mechanism  turned out to have much weaker influence on the transport process than the other connected with the fraction of leaky junctions.  Nevertheless, taking it into account improves the agreement of the computational averaged tissue concentration with the experiments.  

The final conclusion is that proposed model, which takes into account all effects described in this paper, is consistent with experimental evidence. The model is able to quantitatively predict increased LDL accumulation in the arterial wall under hypertension. It has to be stressed, however, that the small resolution of the experimental procedure rather excludes possibility of theprecise and unambiguous validation of the modelling approach. 

%------------------------------------------------
%------------------------------------------------
\section{Acknowledgments}
The study was supported by National Science Centre grant NCN2014/15/B/ST7/05271.
K.J. and A.S. acknowledge a scholarship from the FORSZT project co-financed by the European Social Fund.

%------------------------------------------------
%------------------------------------------------
\section{Literature}
\bibliographystyle{elsarticle-num}
\bibliography{paper}
\end{document}